\documentclass[12pt]{article}

\def\src{SGR\,0418$+$5729}

\def\ea {1E\,2259$+$586}

\def\psr{PSR\,1622$-$4950}
\def\hbpsr{PSR\,J1846$-$0258}

\def\nh {$N_{H}$}

\def\ergs {erg\,s$^{-1}$}
\def\ergscm2 {erg\,s$^{-1}$cm$^{-2}$}
\def\ss {s\,s$^{-1}$}
\def\cm2 {cm$^{-2}$}

\usepackage{graphicx}
\usepackage{url}
\usepackage{color}

\usepackage{scicite}

\usepackage{times}

\newenvironment{sciabstract}{%
\begin{quote} \bf}
{\end{quote}}

\newcounter{lastnote}
\newenvironment{scilastnote}{%
\setcounter{lastnote}{\value{enumiv}}%
\addtocounter{lastnote}{+1}%
\begin{list}%
{\arabic{lastnote}.}
{\setlength{\leftmargin}{.22in}}
{\setlength{\labelsep}{.5em}}}
{\end{list}}

\title{A low-magnetic-field Soft Gamma Repeater}

\author
{N. Rea$^{1\ast}$, P. Esposito$^{2}$, R. Turolla$^{3,4}$, G. L. Israel$^{5}$, \\
S. Zane$^{4}$, L. Stella $^{5}$,  S. Mereghetti$^{6}$, A. Tiengo$^{6}$,\\
  D. G\"otz$^{7}$, E. G{\"o}{\u g}{\"u}{\c s}$^{8}$, C. Kouveliotou$^{9}$
\\
\normalsize{$^{1}$Institut de Ci\'encies de l'Espai (CSIC--IEEC), Facultat de Ci\'encies, Campus UAB,}\\
\normalsize{Torre C5-parell, 2a planta, 08193, Bellaterra (Barcelona), Spain}\\
\normalsize{$^{2}$INAF - OAC, loc. Poggio dei Pini, strada 54, I-09012 Capoterra, Italy.}\\
\normalsize{$^{3}$Dipartimento di Fisica, Universit\`a di Padova, via F. Marzolo 8, I-35131 Padova, Italy.}\\
\normalsize{$^{4}$MSSL-UCL, Holmbury St. Mary, Dorking, Surrey RH5 6NT, UK.}\\
\normalsize{$^{5}$INAF - OAR, via Frascati 33, I-00040 Monteporzio Catone, Italy.}\\
\normalsize{$^{6}$INAF - IASF Milano, via E. Bassini 15, I-20133 Milano, Italy.}\\
\normalsize{$^{7}$AIM (CEA/DSM-CNRS-Universit\'e Paris Diderot), Irfu/Service d'Astrophysique,}\\
\normalsize{Saclay, F-91191 Gif-sur-Yvette, France.}\\
\normalsize{$^{8}$Sabanc\i\ University, Orhanl\i-Tuzla, 34956 \.Istanbul, Turkey.}\\
\normalsize{ $^{9}$NASA Marshall Space Flight Center, Huntsville, AL 35812, USA}\\
\normalsize{$^\ast$To whom correspondence should be addressed; E-mail: rea@ieec.uab.es.}
}

\date{}

\begin{document}


\baselineskip24pt


\maketitle


\begin{sciabstract}
Soft gamma repeaters and anomalous x-ray pulsars form a rapidly
increasing group of x-ray sources exhibiting sporadic emission of
short bursts. They are believed to be magnetars, i.e. neutron
stars powered by extreme magnetic fields, $B\sim10^{14}-10^{15}$
Gauss. We report on a soft gamma repeater with low magnetic field, \src ,
recently detected after it emitted bursts similar to those of
magnetars. X-ray observations show that its dipolar magnetic field
cannot be greater than $7.5\times 10^{12}$\,Gauss, well in the range
of ordinary radio pulsars, implying that a high surface dipolar
magnetic field is not necessarily required for magnetar-like
activity. The magnetar population may thus include objects with a
wider range of B-field strengths, ages and evolutionary
stages than observed so far.

\end{sciabstract}

Magnetized, isolated rotating neutron stars are often detected as pulsating sources in
the radio and x-ray bands, hence the name pulsars. Pulsars slow down with time as their rotational energy is lost via magnetic dipole radiation. The surface dipolar magnetic field (B) of a pulsar can be estimated using its spin period, $P$, and spin-down rate, $\dot{P}$, as follows:

\begin{equation}
B  = (3~I~c^3 \dot{P}~P/8\pi^2 R^6)^{1/2} \sim 3.2\times10^{19}(P \dot{P})^{1/2}~{\rm Gauss}
\label{bspind}
\end{equation}
where $P$ is in seconds, $\dot{P}$ in seconds/second, and we assumed $R\sim10^6$~cm and $I\sim10^{45}$~g~cm$^2$, which are the neutron star radius and moment of inertia, respectively.

Although this expression was developed to estimate the magnetic fields of radio pulsars, usually $\sim10^{12}$ Gauss, it has been traditionally used also for magnetars, where the derived values of $B$ reach $\sim 10^{15}$Gauss (\emph{1}). 
To date only $\sim16$ of these ultra-magnetized neutron stars have been observed  (\emph{\ref{mereghetti08}}, 6); their population includes soft gamma
repeaters (SGRs) and anomalous x-ray pulsars (AXPs). All known magnetars are x-ray pulsars with luminosities of
$L_{\rm X}\sim 10^{32}$--$10^{36}$ \ergs, usually much higher than
the rate at which the star loses its
rotational energy through spin-down. Their high luminosities
together with the lack of evidence for accretion from a stellar
companion (\emph{\ref{mereghetti98}, \ref{dib08}}), led to the
conclusion that the energy reservoir fueling the SGR/AXP activity
is their extreme magnetic field (\emph{\ref{td95}, \ref{td96}}).
Observationally, magnetars are characterized by stochastic
outbursts (lasting from days to years) during which they emit very
short x/$\gamma$-ray bursts; they have rotational periods in a
narrow range (2--12 s) and, compared to other isolated neutron
stars, large period derivatives of $\sim10^{-13}-10^{-10}$ \ss.
Their large dipolar $B-$fields and relatively young characteristic
ages are estimated to be over $\sim 5\times 10^{13}$ Gauss, and
$t_{\rm c} = P/2\dot{P}\sim$0.2 kyr - 0.2 Myr [see
(\emph{\ref{mereghetti08}}) for a review].

In addition to the canonical SGRs and AXPs, two other sources are
known to show magnetar-like activity: \hbpsr\ (\emph{\ref{gavriil08},
  \ref{samar08}}) and \psr\ (\emph{\ref{levin10}}). The former is a 0.3\,s, allegedly rotation-powered, x-ray pulsar, with a
magnetic field of $B\sim4.8\times10^{13}$\,Gauss (in the lower end of the magnetar range), from which a typical magnetar outburst and short x-ray bursts were detected. In the latter, flaring radio
emission with a rather flat spectrum (similar to those observed in
the two transient radio magnetars; (\emph{\ref{camilo06},
\ref{camilo07}})) was detected from a 4.3\,s radio pulsar with a magnetic
field in the magnetar range ($B\sim3\times10^{14}$\,Gauss).

In all sources with magnetar-like activity, the dipolar field
spans $5\times10^{13}\,{\rm G} <B< 2\times10^{15}\,{\rm G}$, which
is $\sim 10$--1000 times the average value in radio pulsars and
higher than the electron quantum field, $B_{\rm
Q}=m_e^2c^3/e\hbar\sim 4.4\times10^{13}$ Gauss. The
existence of radio pulsars with $B > B_{\rm Q}$  and showing only normal
behavior (\emph{\ref{kaspi10}}) is an indication that a magnetic field larger than the quantum electron field alone may not be
a sufficient condition for the onset of magnetar-like activity. In contrast, so far  the opposite always held: magnetar-like activity was observed only
in sources with dipolar magnetic fields stronger than $B_{\rm Q}$.

\src\, was discovered on 5 June 2009 when
the Fermi Gamma-ray Burst Monitor (GBM) observed two
magnetar-like bursts (\emph{\ref{vanderhorst10}}). Follow-up
observations with several x-ray satellites show that it has x-ray
pulsations at $\sim$9.1\,s, well within the range of periods of magnetar
sources (\emph{\ref{gogus09}, \ref{esposito10}}). Further studies
show that \src\ exhibits all the typical characteristics of
a magnetar: i) emission of short x-ray bursts, ii) enhanced
persistent flux, iii) slow pulsations with a variable pulse
profile, and iv) a x-ray spectrum characterized by a thermal plus
non-thermal component, which softened as the outburst decayed.

What made this source distinctly different was the failure of
detecting a period derivative in the first 160 days after the outburst onset, despite frequent observational coverage. Several
x-ray satellites (\emph{\ref{esposito10}}) monitored the source
almost weekly since its detection. This extensive observational
campaign allowed the determination of an accurate ephemeris for
the pulsar rotational period, but no sign of a spin down was detected. In the first 160 days after the outburst onset, the
upper limit on the period derivative was $10^{-13}$ \ss\ (90\%
confidence level), which, according to equation (\ref{bspind}),
translates into a surface dipolar magnetic field $B <
3\times10^{13}$ Gauss (\emph{\ref{esposito10}}). This limit is quite
low for a magnetar source, but not abnormally so, given the
detection of a comparable magnetic
field in the magnetar-like \hbpsr\ (\emph{\ref{gavriil08}}), or the case of AXP \ea\ with $B\sim 6\times
10^{13}$\, Gauss (\emph{\ref{gavriil02}}).

\src\, could not be monitored for a while after the first 160 days,
because the Sun became too close to its position in the sky. On 2010
July 9th, soon after it became observable again, we started an
extensive monitoring of the source with the Swift, Chandra and XMM-Newton
X-ray satellites (Tab.\,S1). In particular, on 2010 July 23rd, we
detected it with the Advanced CCD Imaging Spectrometer (ACIS) onboard
Chandra at a flux of $(1.2\pm0.1)\times10^{-13}$\ergscm2 \,(0.5--10\,keV),
more than one order of magnitude fainter than in the previous
available observation (\emph{\ref{esposito10}}). The spectrum is well
fit by an absorbed blackbody with a line of sight absorption
\nh$=(1.5\pm1.0)\times10^{21}$\cm2 \, and $kT=0.67\pm0.11$\,keV (all
quoted errors are at 90\% confidence level). Pulsations were also
clearly detected at the known magnetar period. On 2010 September 24th,
we observed \src\, with the European Photon Imaging Camera (EPIC) 
onboard XMM--Newton, which detected it at a comparable flux, and
could measure again the rotational period of the neutron star. We
used our new Swift, Chandra and XMM--Newton observations, together
with several other observations (Tab.\,S1) and phase-connected all the
source data from 2009 June 5th till 2010 September 24th (Fig.\,1 and
Supporting on-line Material).  We found a best fit period of
9.07838827(4)\,s referred to TJD (Truncated Julian Day) 14993.0 and
to the Solar System barycenter. The phase evolution of \src\ is well
described by a linear relation $\phi=\phi_0 + 2\pi (t-t_0)/P$, and a
quadratic term $-2\pi \dot P(t-t_0)^2/2 P^2$ (which reflects the
presence of a spin down), is not statistically required. This implies
an upper limit on the period derivative of \src\ of $\dot{P} <
6.0\times10^{-15}$\ss\ (90\% confidence level). This value is the
smallest of all known SGRs/AXPs, of the two magnetar-like pulsars
\hbpsr\ and \psr, and of the X-ray Dim Isolated Neutron Stars (XDINSs;
\emph{\ref{tu09}}) for which a measure of $\dot P$ is available
(Fig.\,2). The corresponding limit on the surface dipolar magnetic
field of \src\, is $B < 7.5\times10^{12}$\,Gauss, making it the
magnetar with the lowest surface dipolar magnetic field yet. The upper
limit on the period derivative implies a characteristic age of the
source $t_{\rm c} > 24$\, Myr.

Despite the characteristic age is known to overestimate the true age of a neutron star in which magnetic field decay occurred (\emph{\ref{co00}}), as is likely the case of \src, the rather high Galactic latitude ($b=5.1$\,deg) and its
position on the $P$--$\dot P$ plane [close to the death line for radio
emission (\emph{\ref{cr80},\ref{zhang00}})], suggest that this system is quite
older than the other SGRs/AXPs. 

The existence of magnetar-like sources with low values of $B$ has
several consequences.  Among isolated pulsars, which are presumably
rotation-powered, $\sim$18\% have a dipolar magnetic field higher than
the upper limit we derived for \src\, (Fig.\,2). The discovery  of
\psr\ (\emph{\ref{levin10}}) on the other hand, suggests that
magnetar-like behavior may manifest itself mostly in the radio band. In
this framework, our result indicates that a large number of apparently
normal pulsars might turn on as magnetars at anytime, regardless of
having a surface dipole magnetic field above the quantum limit or not. As a
direct consequence, magnetar-like activity may occur in pulsars with a very
wide range of magnetic fields and it may fill a continuum in
the $P-\dot{P}$ diagram (Fig.\,2).

So far we have been considering the relationship between the
surface dipolar magnetic field and magnetar-like activity. However,
it is likely that the magnetar activity is driven by the magnetic
energy stored in the internal toroidal field (\emph{\ref{td95}, \ref{td01}});
this component cannot be measured directly. If the
magnetar model as it is currently understood is indeed valid,
despite its low surface dipolar field,  \src\ is expected to harbor a
sufficiently intense internal toroidal component $B_{\rm tor}$ in
order to be able to undergo outbursts and emit bursts. This large
internal field can stress the crust and ultimately deforms/cracks the
star surface layers, periodically allowing magnetic helicity to be
transferred to the external field, thus causing the (repeated) short
x-ray bursts and the overall magnetar-like activity (\emph{\ref{td95},\ref{tlk02},
\ref{belo09}}).

As with other magnetars, $B_{\rm tor}$ can be estimated assuming that
the magnetic energy stored in the internal toroidal field powers the
quiescent emission of \src\ during its entire lifetime, $B_{\rm tor}^2
\sim 6 L_{\rm X} t_{\rm c} / R_{\rm NS}^3$
(\emph{\ref{td95}}). Assuming a source distance of 2\,kpc
(\emph{\ref{vanderhorst10}, \ref{esposito10}}), and that the current
luminosity $L_{\rm X}\sim6.2\times10^{31}$\ergs\ (the lowest measured
so far for this source) corresponds to the quiescent luminosity, we
obtain $B_{\rm tor} \sim 5\times10^{14}$\, Gauss for a neutron star
radius of $R_{\rm NS}= 10^6$\,cm and a source characteristic age of
$t_{\rm c}\sim24$\,Myr. A value of the same order is obtained if the
ratio of the toroidal to poloidal field strength is $\sim 50$, as in
the magneto-thermal evolution scenario (\emph{\ref{po09}}, 24). In this picture, \src\ may possess a high enough internal magnetic field to overcome the crustal yield and give rise to magnetar-like activity despite its
low surface dipolar magnetic field.  However, would the actual measurement of the surface dipolar $B$-field of \src\, turn
out to be much smaller than the present upper limit, this may require to
rethink some of the ingredients at the basis of the magnetar scenario.

\src\ may represent the tip of the iceberg of a large population of old and low-dipolar-field magnetars
that are dissipating the last bits of their internal magnetic energy (\emph{29}).
Indeed, a large fraction of the radio pulsar population may have
magnetar-like internal fields not reflected in their normal dipolar
component.

\begin{quote}
{\bf References and Notes}
\begin{enumerate}

\item We note that the surface dipolar magnetic field of magnetars has been estimated also with several other methods (\emph{\ref{vietri07},\ref{td95},\ref{td01}}); these give values consistent with those derived from the formula in eq.(1) . 

\item \label{vietri07} M. Vietri, L. Stella \& G.L. Israel {\it Astrophys. J.}  {\bf 661}, 1089 (2007).
\item\label{td95} C. Thompson \& R. C. Duncan, R.C. {\it MNRAS} {\bf 275}, 255 (1995).
\item\label{td01} C. Thompson \& R. C. Duncan, R.C.  {\it Astroph. J} {\bf 561}, 980 (2001).
\item \label{mereghetti08} S. Mereghetti, {\it Astronom. Astrophys. Rev.}  {\bf 15}, 225 (2008).
\item See http://www.physics.mcgill.ca/\textasciitilde pulsar/magnetar/main.html for an updated catalogue of SGRs/AXPs .
\item \label{mereghetti98} S. Mereghetti, G. L. Israel \& L. Stella,  {\it Mon. Not. R. Astron. Soc.} {\bf 296}, 689 (1998).
\item \label{dib08} R. Dib, V. M. Kaspi \& F. P. Gavriil,  {\it Astrophys. J.}  {\bf 666}, 1152 (2008).

\item\label{td96} C. Thompson \& R. C. Duncan, R.C. {\it Astrophys. J.}  {\bf 275}, 322 (1996).
\item \label{gavriil08} F. P. Gavriil, M. E. Gonzalez, E. V. Gotthelf, V. M. Kaspi, M. A. Livingstone, P. M. Woods, {\it Science}  {\bf 319}, 1802 (2008).

\item \label{samar08} H. S. Kumar \& S. Safi-Harb {\it Astrophys. J.}  {\bf 678}, L43 (2008).
\item \label{levin10} L. Levin \emph{et al.}, {\it Astrophys. J}, {\bf 721}, L33 (2010).
\item \label{camilo06} F. Camilo, S. M. Ransom, J. P. Halpern, J. Reynolds, D. J. Helfand, N. Zimmerman, J. Sarkissian, {\it Nature}, {\bf 442}, 892 (2006).
\item \label{camilo07} F. Camilo, S. M. Ransom, J. P. Halpern, J. Reynolds, {\it Astroph. J}, {\bf 666}, L93  (2007).
\item \label{kaspi10} V. M. Kaspi, {\it Publ. of the Nat. Academy of Science} {\bf 107}, 7147 (2010).
\item \label{vanderhorst10} A. J. van der Horst {\it et al., Astrophys. J.} {\bf 711}, L1 (2010).
\item \label{gogus09} E. G{\"o}{\u g}{\"u}{\c s}, P. Woods , C. Kouveliotou , \emph{Astron. Telegram}, 2076 (2009).
\item \label{esposito10} P. Esposito \emph{et al.}, {\it Mon. Not. R. Astron. Soc.}  {\bf 405}, 1787 (2010).
\item \label{gavriil02} F. P. Gavriil, V. M. Kaspi, {\it Astroph. J} {\bf 567}, 1067 (2002).
\item\label{tu09} R. Turolla, {\it Astrophysics and
Space Science Library, Neutron stars and pulsars. Springer Berlin}, {\bf 357} (2009).
\item\label{tlk02} C. Thompson, M. Lyutikov, S. R. Kulkarni {\it Astroph. J} {\bf 574}, 332 (2002).
\item\label{belo09} A. M. Beloborodov {\it Astroph. J} {\bf 703}, 1044 (2009).
\item\label{po09}  J. Pons, J. A. Miralles \& U. Geppert  {\it Astronom. Astrophys} {\bf 496}, 207 (2009).

\item Note that the internal field strength required to produce
  crustal cracking should be typically in excess of $10^{14}$\, Gauss
  (\emph{\ref{td95}}).

\item\label{co00} M. Colpi, U. Geppert \& D. Page  {\it Astroph. J} {\bf 529}, L29 (2000)
\item\label{cr80}  A. F. Cheng \& M. A. Ruderman  {\it Astroph. J} {\bf 235}, 576 (1980).
\item\label{zhang00} B. Zhang, A. Harding \& A. G. Muslimov  {\it Astroph. J} {\bf 531}, L135 (2000).
\item \label{manchester05} R. N. Manchester, G. B. Hobbs, A. Teoh, M. Hobbs, {\it Astron. J.}, {\bf 129}, 1993 (2005); the online Australia Telescope National Facility Pulsar Catalogue is available at the web address http://www.atnf.csiro.au/research/pulsar/psrcat.
\item Note that if among these were young, fast spinning young pulsars, the resulting magnetically-induced ellipticity would lead to powerful emission of periodic gravitational waves.
\end{enumerate}
\end{quote}





\begin{scilastnote}

\item NR is supported by a Ram\'on~y~Cajal fellowship through Consejo
  Superior de Investigaciones Cientf\'icas, by grants AYA2009-07391
  and SGR2009-811, and thanks D.~F. Torres for useful discussions. PE
  acknowledges financial support from the Autonomous Region of
  Sardinia through a research grant under the program PO Sardegna FSE
  2007--2013, L.R.  7/2007 ``Promoting scientific research and
  innovation technology in Sardinia''. DG acknowledges the CNES for
  financial support. The work of RT, GLI, LS, SM and AT is partially
  supported by INAF-ASI through grant AAE I/088/06/0. We are grateful
  to H. Tananbaum, N. Gehrels and N. Schartel for granting us Chandra, Swift, and XMM-Newton time, respectively, for this research.

\end{scilastnote}

\clearpage
\begin{figure}
\includegraphics[width=18cm]{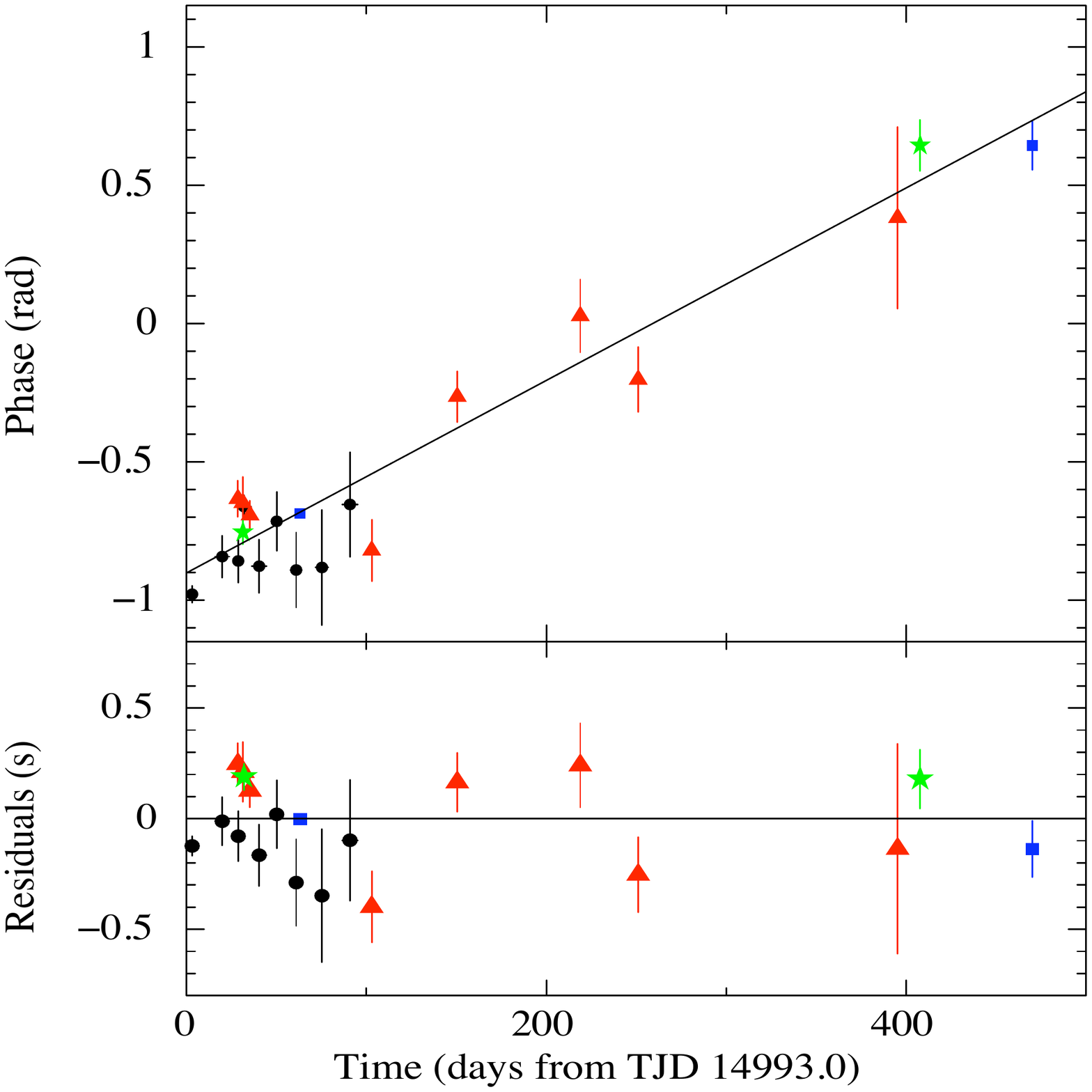}
\end{figure}

\clearpage

\noindent {\bf Fig. 1.} Top panel: rotation phase versus time for the coherent timing solution for \src\, obtained
using data taken with Rossi X-ray Timing Explorer (black circles), Swift (red triangles)
XMM--Newton (blue squares), and Chandra (green stars). The solid line shows the best-fitting linear function ($\chi^2 = 1.8$ for 18 degrees of freedom; root mean square $\sim3$\%). Bottom panel: fit residuals.

\clearpage
\begin{figure}
\includegraphics[width=15cm]{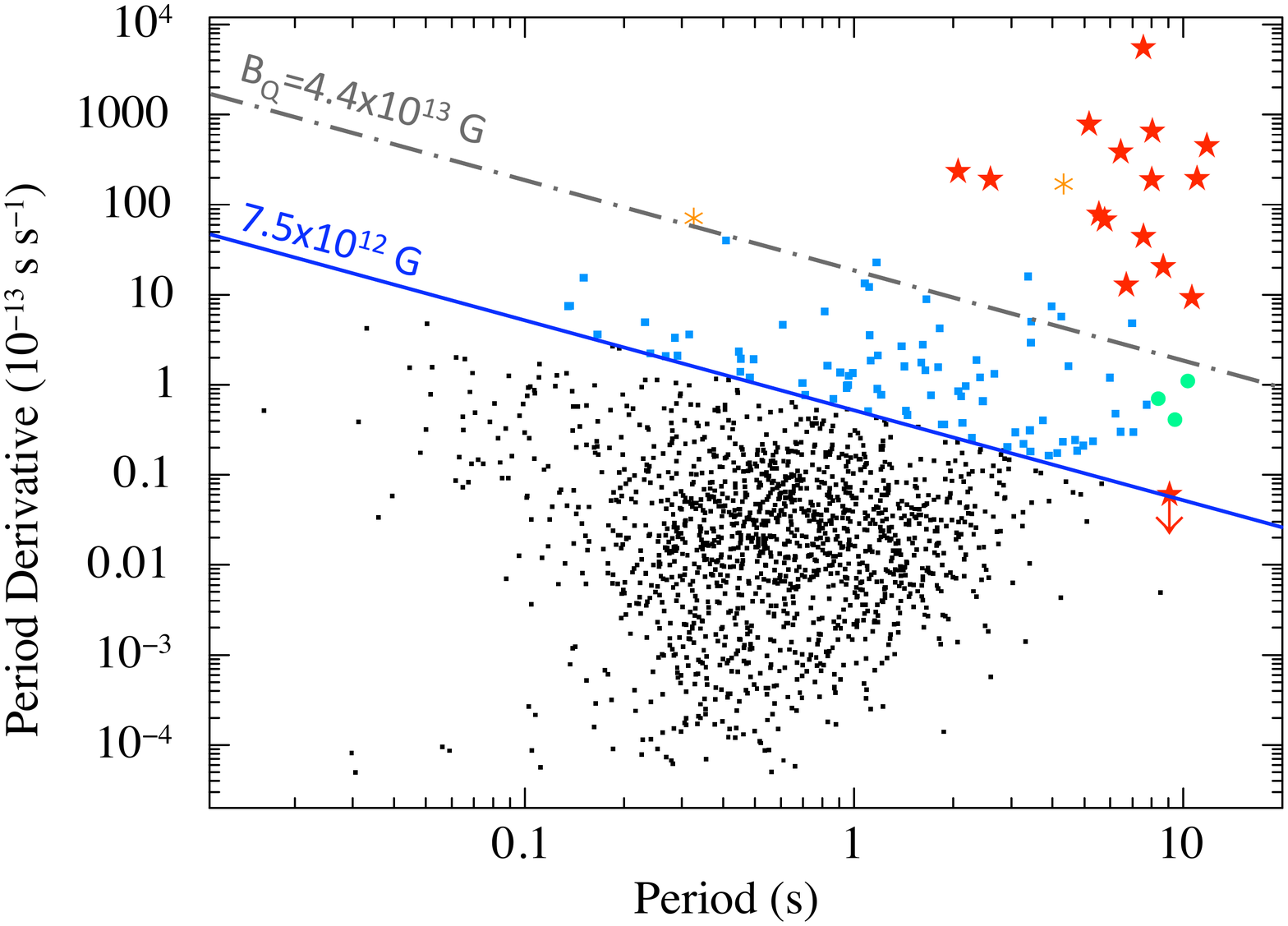}
\end{figure}

\clearpage

\noindent {\bf Fig. 2.}  $P$--$\dot P$ diagram for all known isolated
pulsars [data are from (\emph{\ref{manchester05}})]. $\dot P$ is in
units of $10^{-13}$ \ss .  Black squares represent normal radio
pulsars, light-blue squares normal radio pulsars with a
magnetic field larger than $7.5\times10^{12}$\,Gauss (our limit for \src
), red stars are the magnetars, orange asterisks are the magnetar-like
pulsars \hbpsr\ and \psr, and the green circles are the X-ray
Dim  Isolated Neutron Stars (XDINSs). The blue solid line marks the 90\% upper limit
for the dipolar magnetic field of \src . The value of the electron quantum magnetic field is also reported (dash-dotted
grey line).

\end{document}